\documentclass[prb,twocolumn,showpacs,preprintnumbers,amsmath,amssymb,10pt]{revtex4-1}

\usepackage{color}
\definecolor{blau}{rgb}{0,0,1}

\definecolor{rot}{rgb}{1,0,0}

\usepackage{ulem}

\usepackage[T1]{fontenc}
\usepackage[pdftex]{graphicx}
\usepackage{dcolumn}
\usepackage{bm}

\usepackage{mathcomp}

\DeclareGraphicsExtensions{.pdf}   
\DeclareMathOperator{\e}{e}

\usepackage[justification=raggedright,singlelinecheck=false]{caption}

\begin{document}

\title{Accurate screened exchange LDA band structures for transition metal monoxides MnO, FeO, CoO and NiO}
\author{Roland Gillen}\email{rg403@cam.ac.uk}
\author{John Robertson}
\affiliation{Department of Engineering, University of Cambridge, Cambridge CB3 0FA, United Kingdom}

\date{\today}

\begin{abstract}
We report calculations of the band structures and density of states of the four transition metal monoxides MnO, FeO, CoO and NiO using the hybrid density functional sX-LDA. Late transition metal oxides are prototypical examples of strongly correlated materials, which pose challenges for electronic structure methods. We compare our results with available experimental data and show that our calculations generally yield accurate predictions for the fundamental band gaps and valence bands, in favourable agreement with previously reported theoretical studies. For MnO, the band gaps are still underestimated, suggesting additional many-body effects that are not captured by our screened hybrid functional approach.
\end{abstract}


\maketitle

\section{Introduction}
The oxides of late transition metals are classic examples of strongly correlated systems. While appearing to be simple compounds, the partially filled $d$ shells of the metal ions make them challenging materials for electronic structure theory. 
The exact origin, nature and size of the fundamental gaps of the four oxides has been subject of debate for many years, among both experimentalists\cite{pratt-1959,huffman-1969,powell-1970,bowen-1975,vanderlaan-1981,sawatzky-1984,huefner-1969,vanelp-coo,vanelp-mno,zimmermann-PESBIS} and theorists\cite{terakura-1984,massida-1995,aryasetiawan-1995,faleev-2004,li-2005,patterson-2006,tran-2006,engel-2009,kobayashi-2008,roedl-2009}. 
Despite its successes, density functional theory (DFT) in the framework of the local density approximation (LDA) and the generalized gradient approximation (GGA) is inapplicable to strongly correlated systems. In the case of the MnO and NiO, the predicted band gaps are too small compared to experiments, while CoO and FeO are predicted to be antiferromagnetic but metallic. The reason lies in the incorrect treatment of the exchange interaction in these approximations, which do not sufficiently cancel the electron self-interaction\cite{perdew-sic}. This leads to an underestimation of exchange splitting and the energies of unoccupied states. Clearly, this calls for alternative methods offering a better treatment of exchange effects. The semi-empirical LDA+U\cite{anisimov-1991-lda+u} method modifies LDA by adding an on-site Coulomb interaction (the Hubbard U) to correct the energies of localized semi-core electrons. This method is related to addition of dynamical local correlation effects within dynamic mean field theory (LDA+DMFT), which has been applied to selected transition metals oxides in the past, \textit{e.g.} in Ref.\onlinecite{nekrasov-2012}. However, the choice of the correct U, whether self-consistently or empirically, and the treatment of double-counting is not straight-forward and electrons are not treated on equal footing.

An alternative, non-local, treatment of electron-electron interaction are quasiparticle calculations at the GW level of approximation\cite{Hedin-GW}. These are usually performed by taking the wavefunctions and eigenvalues of a previous DFT calculation as an input and either calculating a one-shot self-energy correction (G$_0$W$_0$), iteratively updating the Green function G (GW$_0$) or updating both Green function G and screened Coulomb interaction W (scGW) until self-consistency is obtained. While these methods have been used on NiO and MnO, the predictions depend on basis set and underlying approximations\cite{massida-1995,aryasetiawan-1995,faleev-2004,li-2005,patterson-2006,jiang-tmo,kobayashi-2008,roedl-2009}, which make it difficult to draw a conclusive picture. Despite their wide success, quasiparticle methods are difficult for a variety of reasons: The energy dependence of the electron-electron interaction makes GW calculations computationally costly and finding a proper way to achieve self-consistency is non-trivial. On the other hand, the perturbative nature of G$_0$W$_0$, while being computationally more favourable, does only permit access to properties related to the electronic band structure, and its results ultimatively depend on the quality of the input wavefunctions and energies. It is thus imperative to find a method that can reasonably describe the electronic structure of strongly correlated materials and that is capable of computationally economic self-consistent calculation of groundstate properties.

Hybrid functionals are an interesting choice, as they incorporate non-local exchange-correlation effects, while maintaining the possibility of variational total energy calculations at moderate computational cost and thus allowing for self-consistent optimization of geometries. They are firmly rooted within DFT via the framework of generalized Kohn-Sham schemes\cite{Seidl-SX}, which allow for explicitly orbital-dependent non-local exchange-correlation potentials. A fraction of Hartree-Fock exchange is mixed into LDA or GGA, which remedies most of the short-comings of the underlying local functional due to the improved (non-local) treatment of the electron exchange interaction. This can be likened to the self-energy in the well-known COHSEX approximation\cite{Hedin-GW}, where the exchange interaction is statically screened by a Coulomb hole.

Hybrid functionals have been shown to be versatile methods for the study of correlated materials, often on par with quasiparticle methods\cite{clark-sx,trani-2010,gillen-ln2o3}, while generally performing great for sp semiconductors and materials of practical interest, such as GaN or ZnO\cite{clark-sx}.
A number of hybrid functional studies on late transition metal oxides has been reported\cite{roedl-2009,bredow-2000,alfredsson-2004,feng-2004,feng-2004-nio,franchini-2005,tran-2006,gerber-2007,marsman-2008,tran-2012,guo-ti203}, with band gaps varying with the fraction of included Hartree-Fock exchange. Most notably, R\"odl \textit{et al.}\cite{roedl-2009} showed that the screened hybrid functional HSE03, which they used as input for their G$_0$W$_0$ calculations, yields good band gaps on its own. We recently reported that the screened hybrid functional 'screened-exchange LDA' (sX-LDA) successfully reproduces the electronic properties of CeO$_2$, as well as those of lanthanide (X=La,Ce,...,Lu)\cite{gillen-ln2o3} and transition metal (X=Ti, Cr, Fe)\cite{guo-ti203} sesquioxides X$_2$O$_3$. Interestingly, we observed that variations in treatment of the screened exchange interaction in the hybrid functional approach can lead to qualitative differences of the predicted properties.
Motivated by this success, we here report the results from our sX-LDA calculations on the transition metal monooxides MnO, FeO, CoO and NiO. 

We confirm that the inclusion of non-local exchange interaction in the functional is sufficient to restore the insulating nature of all four oxides, while the screening leads to overall excellent agreement with experiment, particularly for the valence band density of states and valence band widths, and compares favourably with other theoretical methods.

\section{Method}
The presented results have been calculated by the use of hybrid functionals in the generalized Kohn-Sham (GKS) formalism\cite{Seidl-SX} of density functional theory. Here, the self-energy of an electron in the crystal is described by a linear combination of an orbital-dependent Hartree-Fock (HF) exchange-type term 
and a density-dependent local term. 
In practice, the long-range contribution of the HF exchange can be approximated very well by the long-range contribution of a local functional\cite{hse03}, giving 'screened' hybrid functionals.
This form is favorable for the use in solids due to the slow convergence of the long-range HF contribution if periodic boundary conditions are used and thus allows for more efficient calculations.

In this work, we used the hybrid functional 'screened exchange-LDA' (sX-LDA)\cite{byklein-SX,Seidl-SX} as implemented in the planewave code CASTEP\cite{castep,clark-sx}. It is given by
\begin{equation*}
E_{\mbox{\tiny xc}}^{\mbox{\tiny sX-LDA}}[{\phi}] = E_{\mbox{\tiny x}}^{\mbox{\tiny HF,SR}}[\phi] - E_{\mbox{\tiny x}}^{\mbox{\tiny LDA,SR}}[n] + E_{\mbox{\tiny xc}}^{\mbox{\tiny LDA}}[n]
\end{equation*}
where the Hartree-Fock exchange is screened by a Thomas-Fermi dielectric function, \textit{i.e}
\begin{eqnarray}
\rho_{\mbox{\tiny ij}}(r)&=&\phi_i^*(\mbox{\bf{r}})\phi_j(\mbox{\bf{r}})\nonumber\\
E^{\mbox{\tiny HF,SR}}_{\mbox{\tiny x}}[{\phi}] &\propto& \sum_{\mbox{\tiny i,j}}^{\mbox{\tiny occ}}\iint \frac{\rho_{\mbox{\tiny ij}}(r)\e^{-k_s|\bf{r}-\bf{r}'|}\rho_{\mbox{\tiny ij}}^*(r')}{|\bf{r}-\bf{r}'|}d^3r'd^3r\nonumber\\
E^{\mbox{\tiny LDA,SR}}_{\mbox{\tiny x}}[\rho] &=& F[\gamma]E^{LDA}_{x}\nonumber\\
F[\gamma] &=& 1-\frac{4}{3}\gamma \arctan\left(\frac{2}{\gamma}\right) \nonumber\\
& &-\frac{\gamma^2}{6}\left[1-\left(\frac{\gamma^2}{4}+3\right)\ln\left(1+\frac{4}{\gamma^2}\right)\right],\nonumber
\end{eqnarray}
with $\gamma=k_s/k_F$, where $k_s$ is the inverse screening length of the exact exchange contribution and $k_F$ is the Fermi wave vector.
We use a fixed value of $k_s$=0.76\,bohr$^{-1}$, which works well for $sp$ semiconductors, in all our calculations 
The four transition metal oxides MnO, NiO, FeO and CoO were modelled by a rhombohedral unit cell containing two metal and two oxygen atom, respectively. All calculations converged to the AFII phase, \textit{i.e.} with antiferromagnetic spin ordering along the crystal [111] direction. To confirm that the AFII phase indeed is the global ground state, we also tested the ferromagnetic and the AFI phases and found the AFII spin-ordering to correspond to the lowest total energy.
\begin{table}
\begin{ruledtabular}
\caption{\label{tab:lattice} Lattice constants and bulk moduli of the cubic antiferromagnetic unit cell from sX-LDA calculations.}
\begin{tabular*}{\columnwidth}{@{\extracolsep{\fill}} c c c c c c}
&&MnO&FeO&CoO&NiO\\
\hline
a&sX-LDA&8.67&8.53&8.65&8.47\\
(in \AA)&Exp.\cite{tran-2006}&8.89&8.668&8.534&8.342\\
&$\Delta$&-2.5\%&-1.6$\%$&+1.4\%&+1.5\%\\
\hline
B$_0$&sX-LDA&181.02&207.66&214.92&230.72\\
(in GPa)&Exp.&147\cite{mizutani-1978},153\cite{makino-2000}&174\cite{mizutani-1978}&181\cite{mizutani-1978}&190-220\cite{mizutani-1978}\\
\end{tabular*}
\end{ruledtabular}
\end{table}
We chose to neglect rhombohedral distortions in this work for computational reasons and fitted the total energies to a Birch–Murnaghan equation of state while keeping the cell angles fixed. Table~\ref{tab:lattice} shows the obtained theoretical lattice constants and bulk moduli. 

The atomic cores of Mn, Co and Ni were described by standard normconserving pseudopotentials from the CASTEP database, while we generated a pseudopotential for Fe using the OPIUM\cite{opium1,*opium2} code.
We treat the ($3d$,$4s$,$4p$) states of the four considered transition metals as valence electrons by plane waves with a cutoff energy of 750\,eV. On GGA level, inclusion of semi-core $3s$ and $3p$ electrons in the calculations did not noticeably affect on the electronic properties. A Monkhorst-Pack grid of 4x4x4 points in the Brillouin zone is sufficient to yield total energies converged to 0.005\,eV. We used a grid of 12x12x12 k-points for the calculation of the density of states plots.

\section{Results and Discussion}
\begin{figure*}[tbh]
\centering
\begin{minipage}{0.47\textwidth}
\includegraphics*[width=\textwidth]{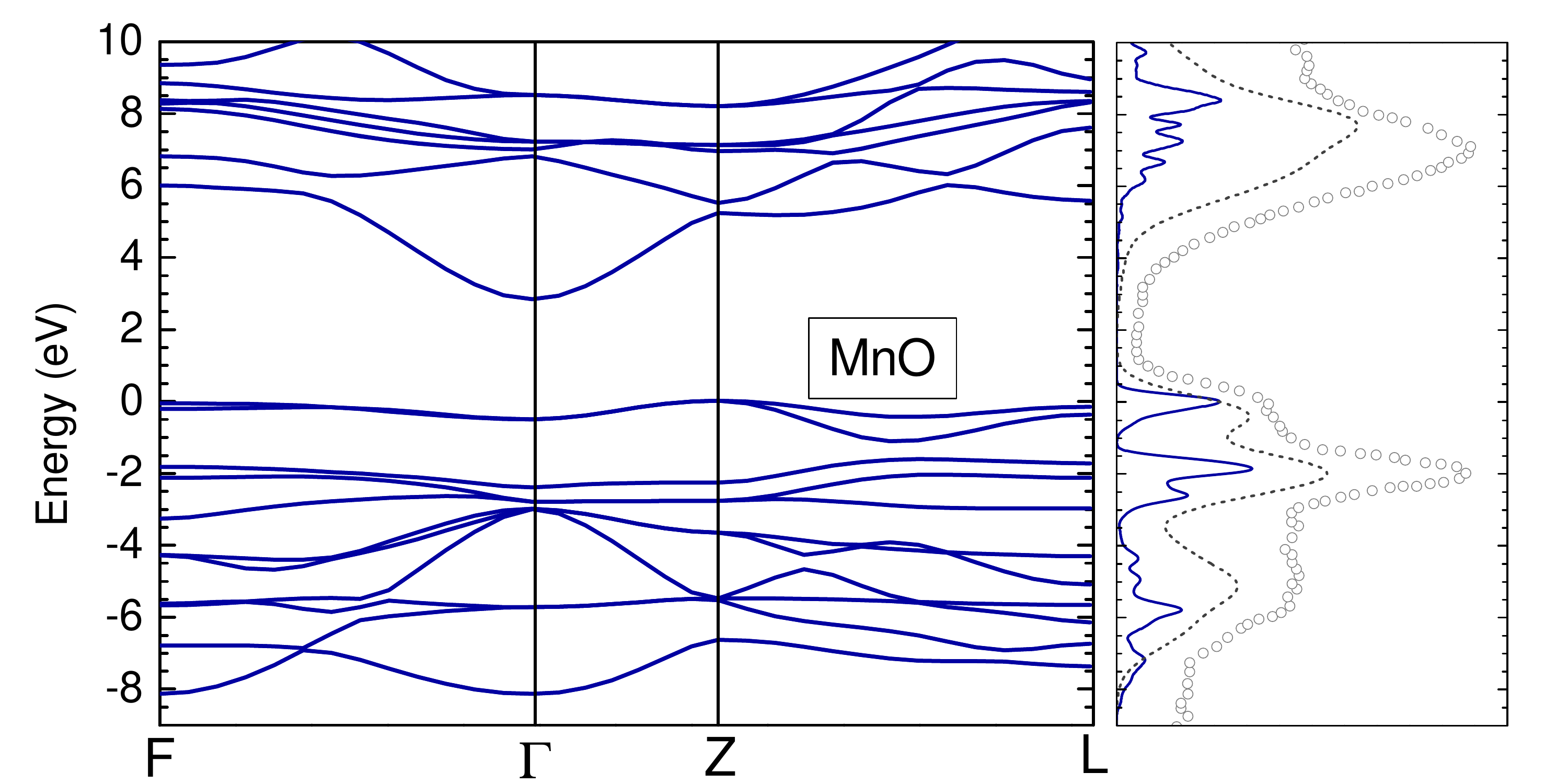}
\end{minipage}
\quad
\begin{minipage}{0.47\textwidth}
\includegraphics*[width=\textwidth]{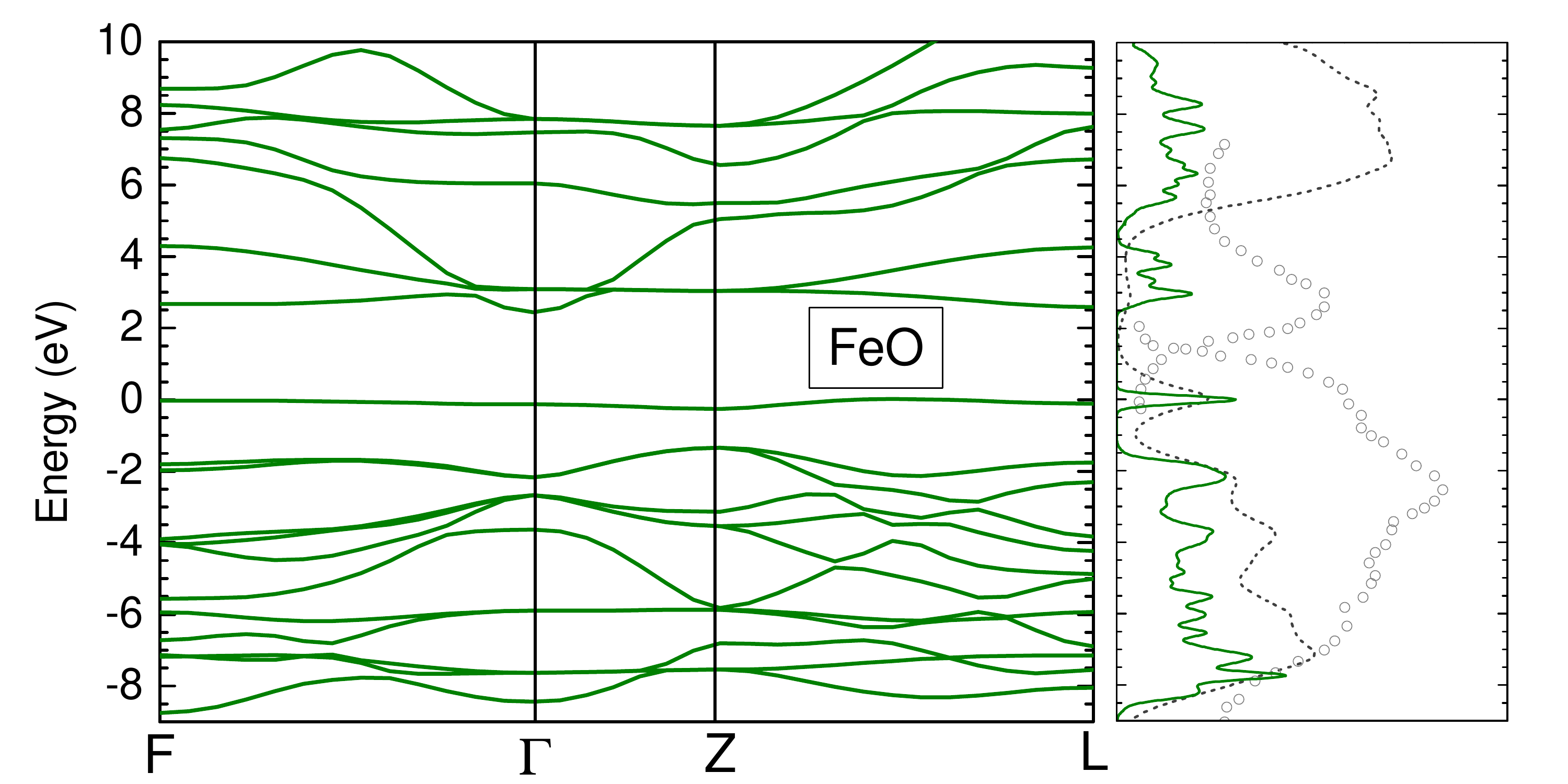}
\end{minipage}
\\
\begin{minipage}{0.47\textwidth}
\includegraphics*[width=\textwidth]{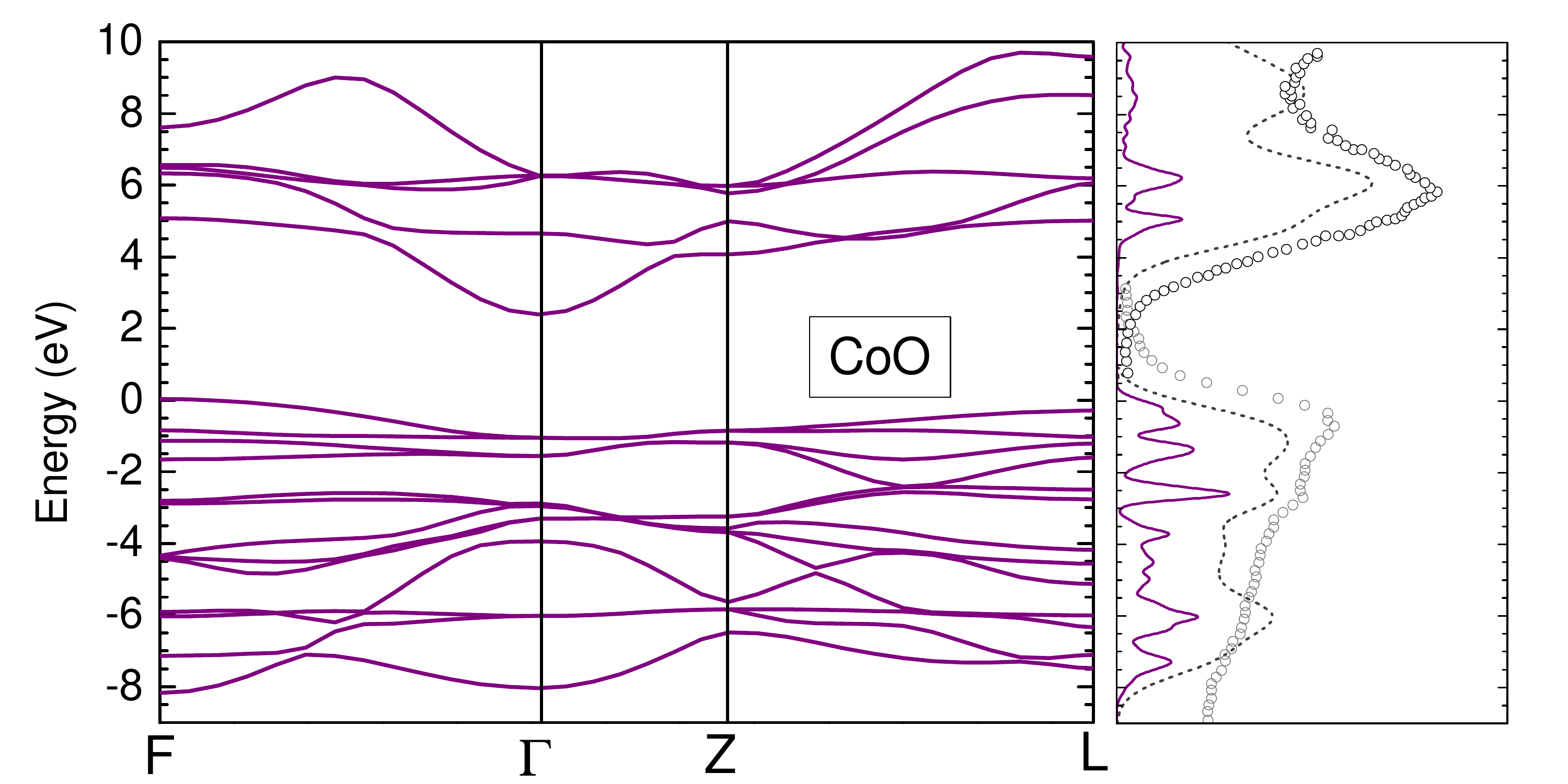}
\end{minipage}
\quad
\begin{minipage}{0.47\textwidth}
\includegraphics*[width=\textwidth]{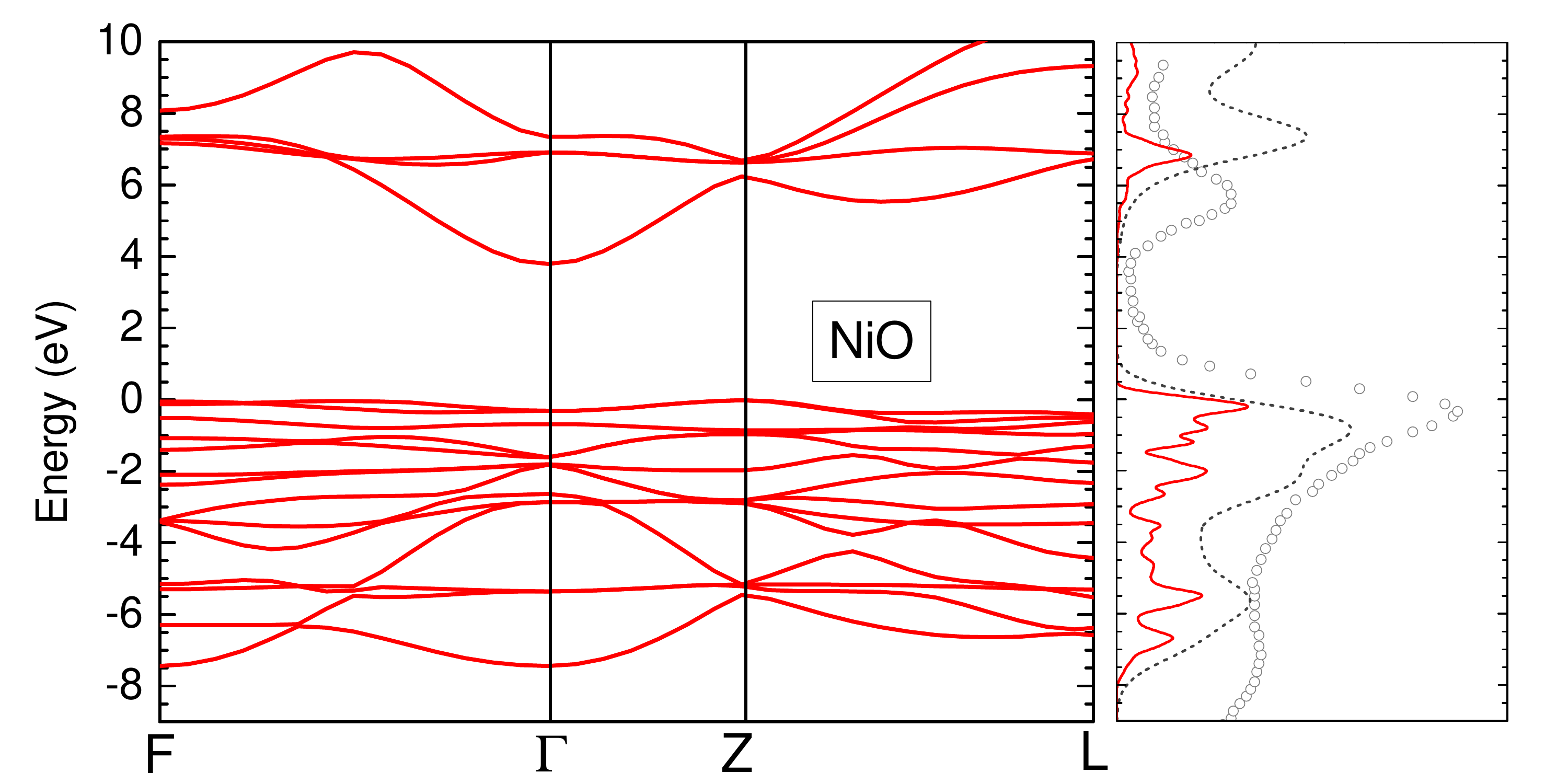}
\end{minipage}
\caption{\label{fig:TMO-bands} (Color online) Calculated electronic band structures and corresponding density of states (coloured lines) and cross-section weighted total density of states (dashed grey line) for MnO, FeO, CoO and NiO. The experimental XPS and BIS spectra (grey circles) are taken from Refs.~\onlinecite{zimmermann-PESBIS} and~\onlinecite{vanelp-coo}. The cross-sectional DOS was calculated by summing the angular momentum channels weighted by their cross-sections\cite{cross2} for the Al K$\alpha$ line and using a Gaussian broadening of the peaks of 0.5\,eV (0.1\,eV for the non-weighted DOS) for better comparability with the XPS and BIS spectra. The energy zero of the experimental spectra was aligned to the calculated Fermi energy.
}
\end{figure*}
\begin{figure*}[tbh]
\centering
\begin{minipage}{0.48\textwidth}
\includegraphics*[width=\textwidth]{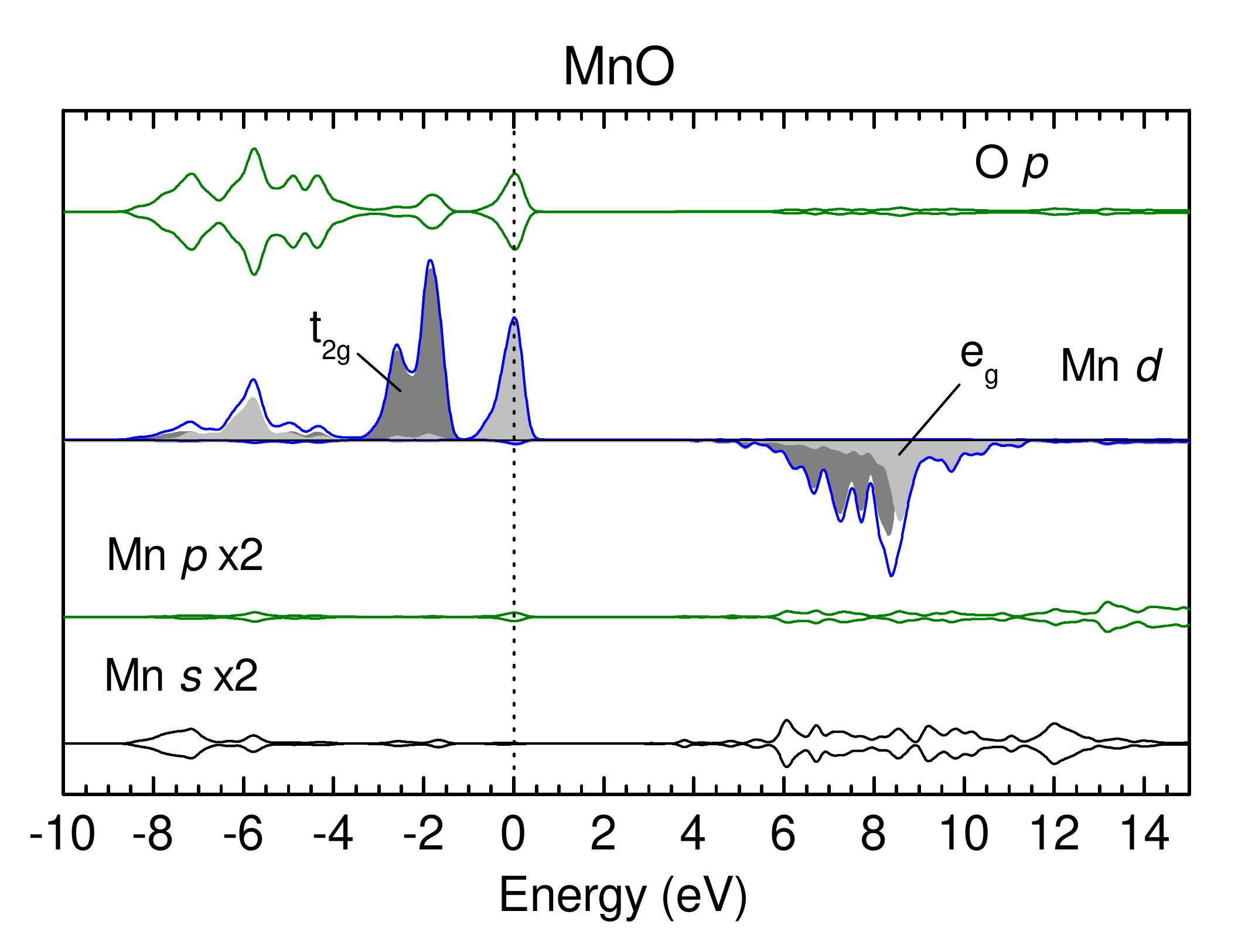}
\end{minipage}
\space
\begin{minipage}{0.48\textwidth}
\includegraphics*[width=\textwidth]{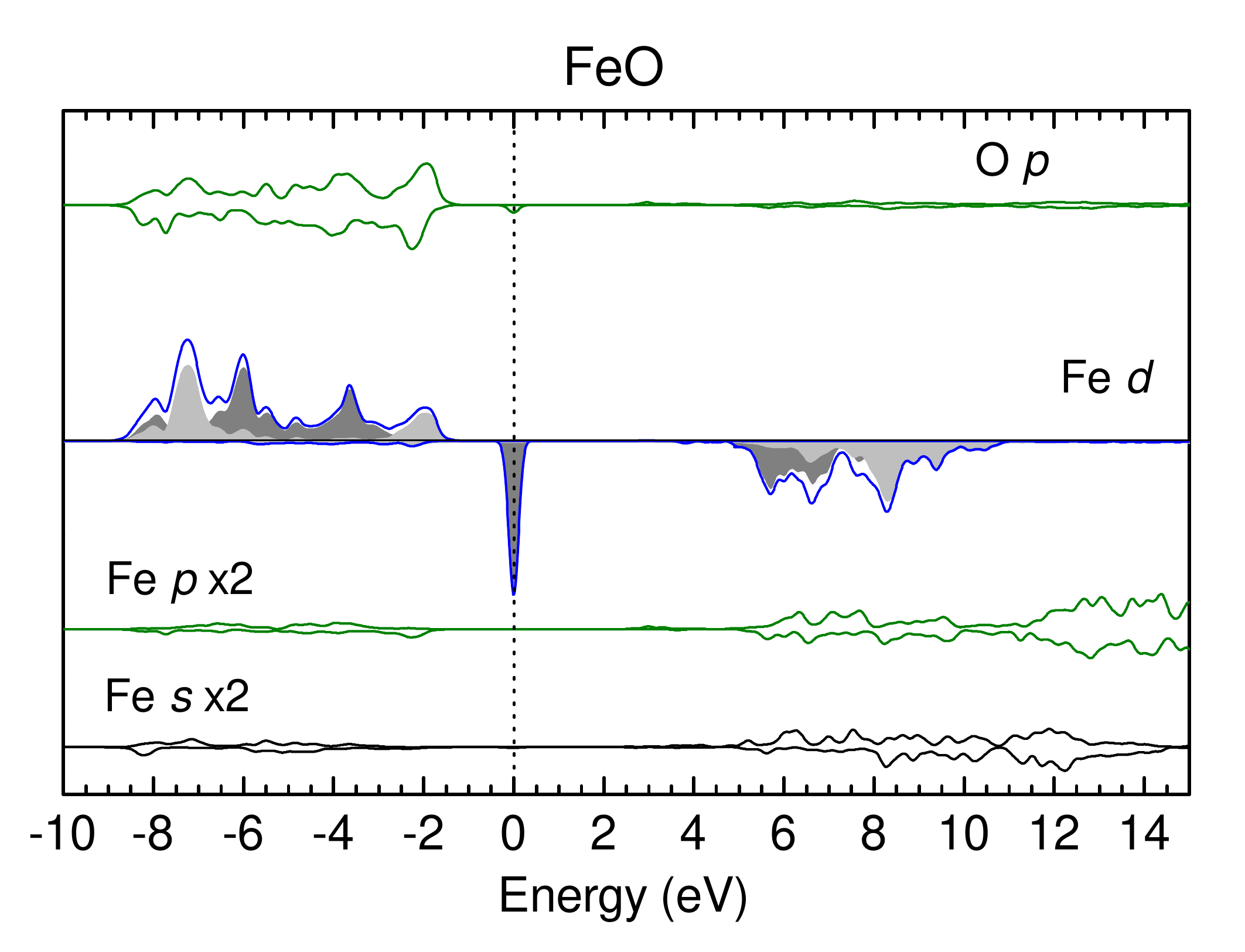}
\end{minipage}
\\
\begin{minipage}{0.48\textwidth}
\includegraphics*[width=\textwidth]{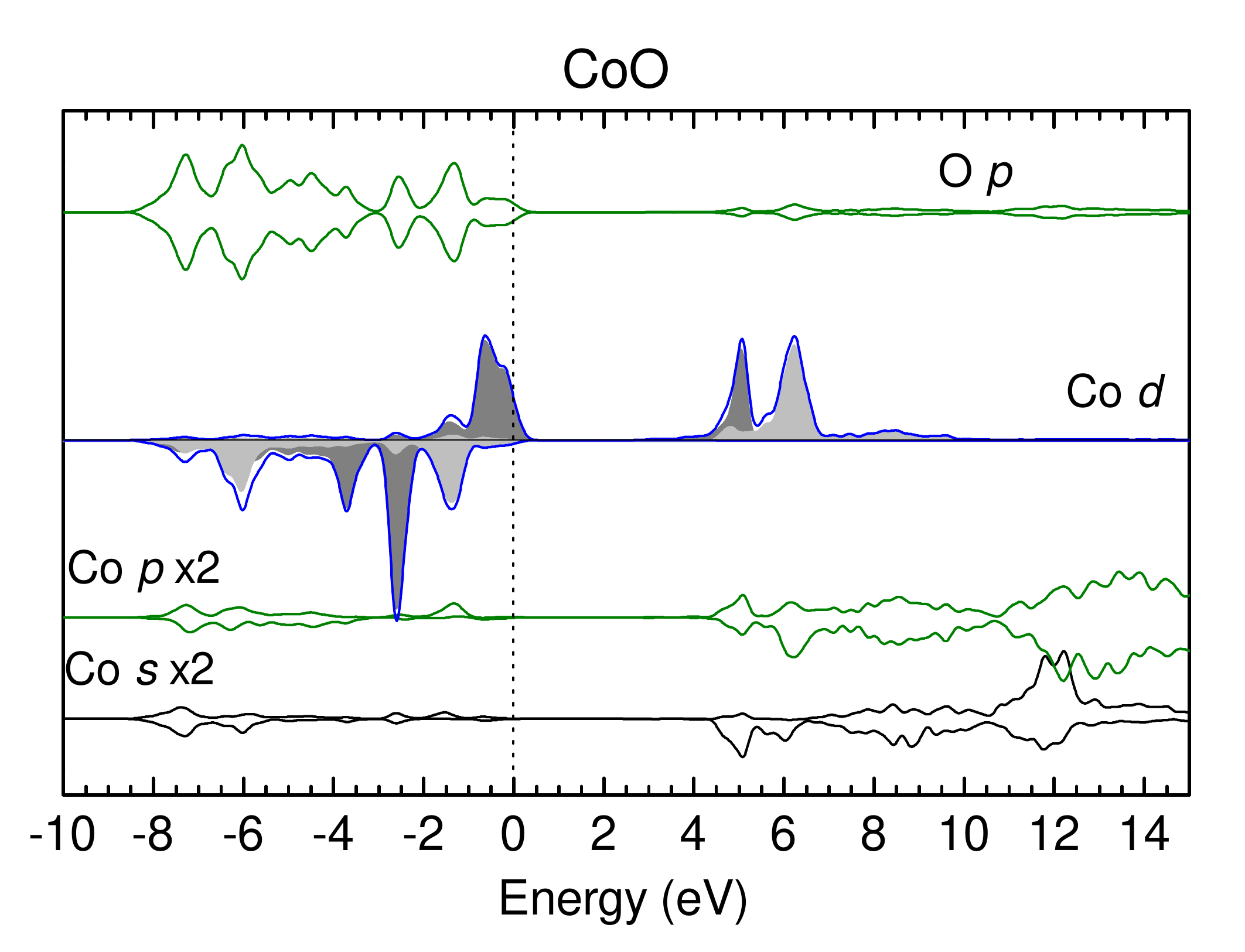}
\end{minipage}
\space
\begin{minipage}{0.48\textwidth}
\includegraphics*[width=\textwidth]{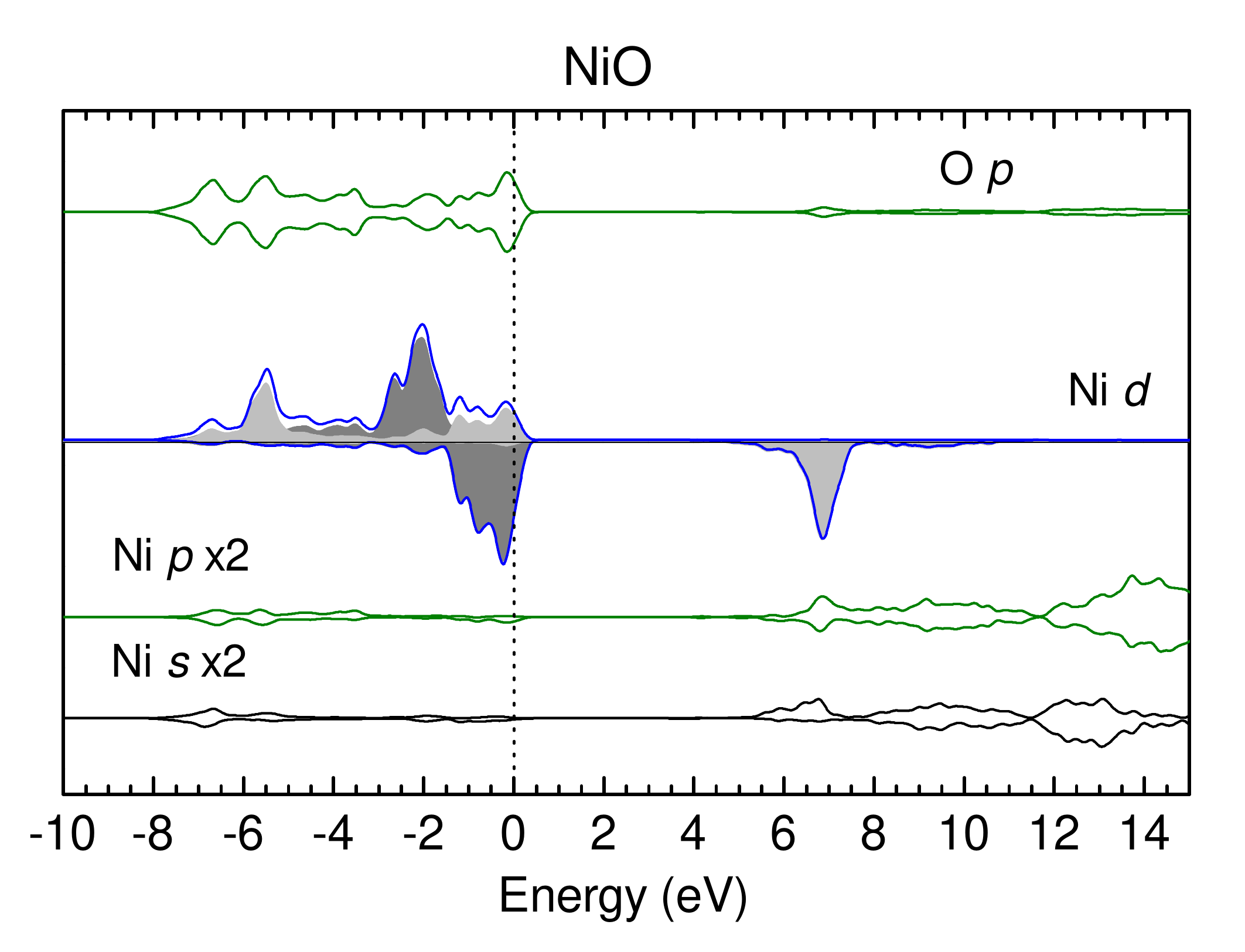}
\end{minipage}
\caption{\label{fig:TMO-pdos} (Color online) Angular momentum channel- and spin-resolved local density of states for MnO, FeO, CoO and NiO from sX-LDA calculations (solid lines). The $d$ states were decomposed into contributions from $e_g$ (light grey area) and $t_{2g}$ (dark grey area)orbitals. The peaks were broadened by a Gaussian of width 0.1\,eV.
}
\end{figure*}

%
Figure~\ref{fig:TMO-bands} shows the band structures from our calculations using the hybrid functional sX-LDA, together with the calculated density of states and reported x-ray photoemission spectroscopy-bremsstrahlung isochromat spectroscopy (XPS-BIS) spectra\cite{zimmermann-PESBIS,vanelp-coo}. In accordance with the previous reports, all of our investigated materials are predicted to be insulating in the antiferromagnetic phase. In all four cases, the valence band top consists of weakly dispersive metal $d$ states with some O $2p$ mixed in. The octahedral crystal field leads to a splitting of the five-fold degenerate $d$ orbitals into two degenerate $e_g$ orbitals and three-fold degenerate $t_{2g}$ orbitals. For MnO, all five $d$ spin-up orbitals are filled and the top $d$ band is of $e_g$ character. Along the series, $d$ orbitals of $t_{2g}$ are successively occupied with minority spin electrons and are shifted down to the valence band top, as seen in the orbital-resolved LDOS in Fig.~\ref{fig:TMO-pdos}. For NiO, only two $d$ bands of $e_g$ remain in the conduction band.
The valence band maximum for MnO and NiO is located at the $Z$ point, whereas we find it between the $Z$ and the $L$ point and at the $F$ point in case of FeO and CoO, respectively. The conduction band minimum for all oxides is a parabolic band of metal $4s$ states, which is centered at $\Gamma$. Correspondingly, the minimum direct band gap is at the $\Gamma$ point.
\begin{table*}
\begin{ruledtabular}
\caption{\label{tab:gaps} Minimum direct and indirect band gaps (in eV) of four transition metal oxides as obtained from different theoretical and experimental methods.}
\begin{tabular*}{\columnwidth}{@{\extracolsep{\fill}} c  c c c c c c c c}
Method$\backslash$Compound&\multicolumn{2}{c}{MnO}&\multicolumn{2}{c}{FeO}&\multicolumn{2}{c}{CoO}&\multicolumn{2}{c}{NiO}\\
&indir&dir&indir&dir&indir&dir&indir&dir\\
\hline
PBE&0.8&1.1&-&-&-&-&0.7&0.9\\
sX-LDA&2.8&3.3&2.45&2.67&2.4&3.4&3.85&4.1\\
sX-LDA (exp. lat. const.)&2.5&3.0&2.3&2.4&2.7&3.7&4.04&4.3\\
HSE03\cite{roedl-2009}&2.6&3.2&2.1&2.2&3.2&4.0&4.1&4.5\\
B3LYP&\multicolumn{2}{c}{3.92\cite{feng-2004}}&3.70\cite{alfredsson-2004}&3.73\cite{alfredsson-2004}&\multicolumn{2}{c}{3.5\cite{bredow-2000}, 3.63\cite{feng-2004}}&\multicolumn{2}{c}{4.2\cite{bredow-2000,feng-2004-nio}}\\
EXX-OEP\cite{engel-2009}&3.85&4.25&1.66&1.7&2.66&3.5&4.1&4.5\\
G$_0$W$_0$@LDA+U&2.34\cite{jiang-tmo},3.05\cite{kobayashi-2008}&3.51\cite{kobayashi-2008}&0.95\cite{jiang-tmo}&&2.47\cite{jiang-tmo}&&3.75\cite{jiang-tmo}, 3.46\cite{kobayashi-2008}&3.97\cite{kobayashi-2008}\\
G$_0$W$_0$@HSE03\cite{roedl-2009}&3.4&4.0&2.2&2.3&3.4&4.5&4.7&5.2\\
Exp (conductivity)&\multicolumn{2}{c}{3.8-4.2\cite{drabkin-1968}}&\multicolumn{2}{c}{}&\multicolumn{2}{c}{3.6\cite{gvishi-1972}}&\multicolumn{2}{c}{3.7\cite{ksendzov-1965}}\\
Exp (XAS-XES)&\multicolumn{2}{c}{4.1\cite{kurmaev-2008}}&\multicolumn{2}{c}{}&\multicolumn{2}{c}{2.6\cite{kurmaev-2008}}&\multicolumn{2}{c}{4.0\cite{kurmaev-2008}}\\
Exp (PES-BIS)&\multicolumn{2}{c}{3.9\cite{vanelp-mno}}&\multicolumn{2}{c}{}&\multicolumn{2}{c}{2.5\cite{vanelp-coo}}&\multicolumn{2}{c}{4.3\cite{sawatzky-1984}}\\
Exp (absorption)&\multicolumn{2}{c}{2.0\cite{huffman-1969},3.6-3.8\cite{iskenderov-1968}}&\multicolumn{2}{c}{2.4\cite{bowen-1975}}&\multicolumn{2}{c}{2.8\cite{pratt-1959}}&\multicolumn{2}{c}{4.0\cite{pratt-1959}}\\
Exp (reflectance)&\multicolumn{2}{c}{}&\multicolumn{2}{c}{}&\multicolumn{2}{c}{2.7\cite{powell-1970}, 5.4\cite{kang-2007}}&\multicolumn{2}{c}{3.7\cite{powell-1970}, 3.9\cite{kang-2007}}\\
\end{tabular*}
\end{ruledtabular}
\end{table*}
\begin{table}
\begin{ruledtabular}
\caption{\label{tab:splitting} Minimum splitting (in eV) of the occupied and unoccupied d-levels for different theoretical methods. We derived the values for GW$_0$@LDA+U from the DOS plots for U=5.4\,eV}
\begin{tabular*}{\columnwidth}{@{\extracolsep{\fill}} c  c c c c}
&MnO&FeO&CoO&NiO\\
\hline
sX-LDA&5&2.7&4.1&5.5\\
sX-LDA (exp. lat. const.)&4.75&2.6&4.4&5.9\\
HSE03\cite{roedl-2009}&~4.5&~3&~4&~4.1\\
B3LYP&4.7\cite{feng-2004}&3.7\cite{alfredsson-2004}&3.4\cite{feng-2004}&4.2\cite{feng-2004-nio}\\
EXX-OEP\cite{engel-2009}&~6&~4&~4&~6.5\\
GW@GGA\cite{li-2005}&&&&4.2\\
scGW\cite{faleev-2004}&6&&&5\\
GW$_0$@LDA+U\cite{jiang-tmo}&~5.75&~3.5&~3.75&~4.25\\
G$_0$W$_0$@HSE03\cite{roedl-2009}&5.5-6&4.5&3.8&5-5.5\\
\end{tabular*}
\end{ruledtabular}
\end{table}
Table~\ref{tab:gaps} shows the band gaps compared to those in previous calculations and from various experimental methods. For experimental lattice constants, our sX-LDA results are quite similar to those given by R\"odl \textit{et al.} for HSE03, even though the difference between indirect and direct band gaps is lower in our sX-LDA calculations. We will discuss our results for the different oxides in comparison with other theoretical methods and experiments in the following.

\subsubsection{FeO}
Unfortunately, the experimental preparation of pure FeO samples is difficult due to Fe segregation\cite{bowen-1975}, which 
hampers comparison of theoretical calculations with experimental spectra. This most likely causes the bad agreement of the experimental XPS-BIS spectrum with theoretical density-of-states plots, see Fig.~\ref{fig:TMO-bands}, which is compatible with the reports from other groups\cite{alfredsson-2004,roedl-2009,sharma-2012}. The bad experimental situation also raises the question of the accuracy of the one reported experimental estimate for the fundamental band gap of FeO\cite{bowen-1975}, 2.4\,eV from optical absorption measurements.

From the theoretical point of view, the fundamental band gap is between a single band of t$_{2g}$ character, which is detached from a block of hybridized O $2p$ and Fe $3d$ bands at lower energies, and the parabolic Fe $4s$ band. Our calculations predict the energy difference between the single $d$ band and the top of the 'bulk' valence band to be approximately 1.1\,eV. The (indirect) fundamental band gap of 2.45\,eV is in slightly better agreement with the experimental value than those from HSE03 and G$_0$W$_0$@HSE03 calculations\cite{roedl-2009}. We believe that this good agreement together with the reported weak quasiparticle correction to HSE03 suggests that the experimental band gap value is indeed of the found magnitude.

Engel \textit{et al.}\cite{engel-2009} recently reported the band structure of FeO from optimized effective potential (EXX-OEP) calculations. While their band structures are qualitatively similar to ours, there are three noticeable differences: In their calculations, the energy difference between the top of the 'bulk' valence band and the conduction band minimum, 4.6\,eV, is about 0.8\,eV larger than in our results. At the same time, the energy of the single $d$ band is predicted to be about 0.75\,eV closer to the conduction bands than in our calculations, closing the fundamental band gap to 1.7\,eV. The third difference is the energy of the unoccupied $d$ bands. Our calculations predict a $d$ band very close to the conduction band minimum, while EXX-OEP shifts them to about 4\,eV above the valence band. A low energy of the unoccupied $d$ states is also suggested by the BIS spectrum in Ref.~\onlinecite{zimmermann-PESBIS}.

\subsubsection{CoO}
For CoO, most experimental reports point towards a band gap size of 2.5-2.8\,eV, but several other studies\cite{gvishi-1972,kang-2007} found significantly higher values. Kang \textit{et al.}\cite{kang-2007} reported a band gap of 5.43\,eV, which they obtained by a 'standard critical point' (SCP) fitting procedure to their ellipsometry spectra. However, they also report a significant optical structure at 2.72\,eV, which they attribute to intra-atomic $d$-$d$ transitions. A Tauc plot of the absorption spectrum we calculated from their measured dielectric function yields an indirect band gap of 2.8\,eV and a direct band gap of $\sim$5\,eV.

Our calculated cross-section weighted density of states in Fig.~\ref{fig:TMO-bands} shows very good agreement with the reported XPS and BIS spectra for CoO\cite{zimmermann-PESBIS,vanelp-coo}. The band structure calculations suggest a fundamental band gap of 2.4\,eV, between a valence band top of hybridized Co $3d$ and O $2p$ states and the minimum of the parabolic band at $\Gamma$. 
This value is comparable to those reported for recent EXX-OEP and G$_0$W$_0$@LDA+U calculations and is close to the range of experimental band gap values. 

The HSE03 and G$_0$W$_0$@HSE03 study by Roedl \textit{et al.}\cite{roedl-2009} yields considerably higher band gaps of magnitude 3.2\,eV and 3.4\,eV, respectively. This is surprising considering the overall close agreement of the results from sX-LDA and HSE03 for the other three oxides, in case experimental lattice constants are used. We thus performed our own HSE03 calculations to test whether the observed divergence is caused by differences in the Co pseudopotentials. Our calculated indirect band gap of 3.15\,eV indeed is in good agreement with the value from Ref.~\onlinecite{roedl-2009}, we thus rule out effects from the pseudopotentials and attribute the differences to details in the structure of the two hybrid functionals. A detailed investigation might be subject of later work.

Finally, we note that Engel \textit{et al.}\cite{engel-2009} found a significant energy gap of approximately 2\,eV between the $d$ bands at the valence band top and the rest of the valence band, similar to the case of FeO. However, we cannot find traces of such a gap neither in the reported photoemission spectra\cite{sawatzky-1984,zimmermann-PESBIS}, nor in other theoretical work on CoO\cite{feng-2004,tran-2006,roedl-2009}.
%
%
\begin{figure}[tbh]
\centering
\includegraphics*[width=0.98\columnwidth]{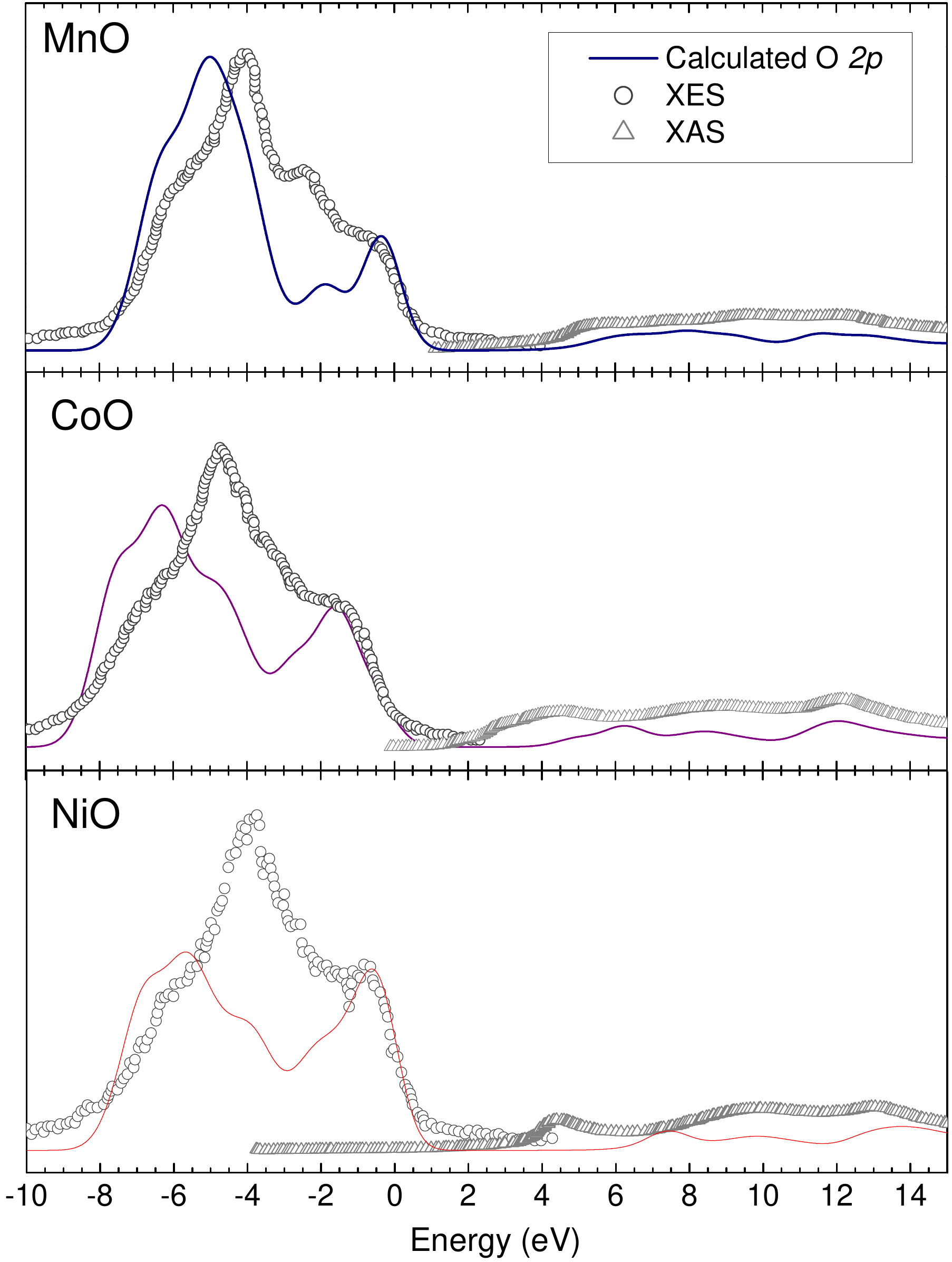}
\caption{\label{fig:TMO-Ostates} (Color online) Comparison of the calculated energy density of oxygen 2p states of MnO (blue solid line), CoO (purple solid line) and NiO (red solid line) with XES-XAS measurements (grey circles and triangles) taken from Ref.~\onlinecite{kurmaev-2008}. The experimental XES and XAS contributions were scaled by two factors to match the peak heights of our calculations. As before, we used a Gaussian broadening of 0.5\,eV.
}
\end{figure}
\subsubsection{MnO and NiO}
Compared to FeO and CoO, MnO and NiO have been investigated more extensively. 
LDA and GGA predict both materials to be insulators, where the band gap of MnO arises from exchange splitting of the $d$ bands and the band gap of NiO arises from a combination of exchange splitting and additional crystal-field splitting between $e_g$ and $t_{2g}$ states. 
The fundamental band gap of NiO was experimentally established to be 3.7-4.3\,eV\cite{ksendzov-1965,sawatzky-1984,pratt-1959,powell-1970,kang-2007} and attributed to transitions from O $2p$ or Ni $3d$ states at the valence band top to either unoccupied Ni $3d$ or Ni $4s$ orbitals\cite{powell-1970}. 
In our case, the final state of the fundamental transition of 3.85\,eV is clearly in the parabolic Ni 4s band, while the onset of the unoccupied $d$ bands is pushed up to 5.9\,eV above the valence band maximum (VBM), see the band structure in Fig.~\ref{fig:TMO-bands}. The parabolic band is of pure Ni $4s$ character at the $\Gamma$-point and of mixed Ni $4s$ and $4p$ character with a minor contribution from O $2p$ away from $\Gamma$. The valence band top arises from a strong hybridization of oxygen $2p$ and nickel $3d$ states, see Fig.~\ref{fig:TMO-pdos}, and is in good agreement with the widely accepted cluster model\cite{asada-1976,*larsson-1976} for the electronic structure of NiO and previous studies on GW and hybrid functional level. We find a spectral weight of 16\% oxygen states and 84\% nickel $3d$ states for the bands within VBM-2\,eV. These values are comparable to the results from GW@GGA calculations by Li \textit{et al.}\cite{li-2005}, who reported a 20\% contribution of oxygen states at the valence band top at $\Gamma$. Our results would thus support both proposed models for the fundamental transition of NiO. For the conduction band, a balanced description of the itinerant "$4s$" bands and the unoccupied $3d$ bands is necessary.

Independent of the method, the inclusion of non-local exchange interaction generally leads to an improved splitting of the occupied and unoccupied $d$ bands in NiO compared to LDA/GGA. On the other hand, the "correction" of the itinerant "$4s$" bands seems to be strongly influenced by the underlying wavefunctions. Li \textit{et al}\cite{li-2005} reported GW@GGA calculations, where the unoccupied $d$ bands are shifted to an energy of $\sim$4.2\,eV above the valence band maximum, while the parabolic band was only weakly shifted compared to GGA. The corresponding DOS of the unoccupied levels is in excellent agreement with BIS\cite{zimmermann-PESBIS} spectra and can account for both the dominant peak at 4.3\,eV and the onset of strong optical absorption and at $\sim$3\,eV. In contrast, the scGW calculations by Faleev \textit{et al.}\cite{faleev-2004}, which feature fully selfconsistent wavefunctions, predicted a much smaller energy difference between the minima of the itinerant band (VBM+4.8\,eV) and the $3d$ states (VBM+5\,eV), with both being at considerably higher energies than in GW@GGA. 
Faleev \textit{et al.}\cite{faleev-2004} attributed their results to underestimated screening from neglecting higher order correlation effects in their calculations, the favourable results from Li \textit{et al.}\cite{li-2005} might thus benefit from error cancellation between the RPA level correlation and the GGA wavefunctions. 

Using HSE03 wavefunctions and eigenvalues for a G$_0$W$_0$ one-shot calculation seems to have a similar effect as the fully self-consistent scheme of Faleev \textit{et al}\cite{faleev-2004} and leads to an overestimated indirect band gap of 4.7\,eV. In contrast, G($_0$)W$_0$ on top of LDA+U wavefunctions and pure HSE03 calculations predict considerably smaller gaps of 3.75\,eV and 4.1\,eV, respectively. Our obtained value of 3.85\,eV falls right in the middle and is in excellent agreement with the experimental values. We note that the use of the slightly smaller experimental lattice constants leads to a larger band gap of 4.04\,eV, which is very close to the HSE03 prediction for the same cell volume.

On the other hand, the splitting of unoccupied and occupied $d$ states in our sX-LDA is even more pronounced than those reported from GW calculations and clearly too strong if one compares the calculated DOS with experimental XPS and BIS spectra for NiO (Fig.~\ref{fig:TMO-bands}) and the HSE03 calculations. 

We note that the strong shift of unfilled semi-core states compared to HSE03 and HSE06 is a known behaviour of sX-LDA\cite{gillen-TCO,gillen-ln2o3} and is likely caused by the different inclusion of non-local exchange. While both functional types are conceptually similar in the sense that they employ a range separation scheme to screen the non-local exchange and generally yield similar band gaps, sX-LDA uses 100\% of Thomas-Fermi screened Hartree-Fock exchange, whereas HSE only incorporates only 25\% Hartree-Fock exchange, but with Error function screening and a weaker screening length compared to sX-LDA. The very strong contribution of Hartree-Fock exchange in short-range within sX-LDA particularly affects more localized orbitals, such as semi-core $d$ electrons, and often leads to strong renormalizations of the corresponding energy levels. On the other hand, the typically stronger screening of the non-local exchange contribution in sX-LDA compared to HSE often leads to a weaker shift of the itinerant states in the conduction band, \textit{e.g.} the "4s" states in case of the transition metal monooxides shown in this work, the Cu p states in copper-based transparent conducting oxides\cite{gillen-TCO}, or the antibonding sp$^3$ states in silicon.

In contrast, HSE03 predicts the $d$ states at significantly lower energies. As a result, the minima of both the "$4s$" band and the "$3d$" band are at about 4.1-4.2\,eV, so that $3d\rightarrow 4s$, $3d\rightarrow 3d$ and $2p\rightarrow 3d$ transitions are candidates for the fundamental transition. In this sense, the predictions of sX-LDA are a quantitative mixture of the 4s band from HSE03 and the strong shift of the $d$ states from quasiparticle methods.

We find a similar 'intermediate' prediction between HSE03 and G$_0$W$_0$ for the case of MnO. HSE03 and sX-LDA both yield fundamental band gaps, which are considerably smaller than the experimental values of 3.6-4\,eV, see Table~\ref{tab:gaps}. We obtain a fundamental band gap of 2.8\,eV, which is somewhat larger than the reported HSE03 value of 2.6\,eV. However, we note that the prediction from sX-LDA is very close to the HSE03 gap if the experimental lattice constants are used.
The quasiparticle corrections R\"odl \textit{et al.} found for the Mn $4s$ band in MnO are quite high (0.8\,eV). This might indicate that the contribution of middle-range Hartree-Fock exchange in sX-LDA and HSE, which mainly leads to a rigid shift of the unoccupied states, is not strong enough in the case of MnO.
An indicator is the fact that the splitting of occupied and unoccupied $d$ bands in the hybrid functional calculations is considerably weaker than in GW, even though the relative energies of the minima of the $4s$ band and the $d$ bands is 2-2.5\,eV in all methods, see Table~\ref{tab:splitting}, compared to the indirect band gaps in Table~\ref{tab:gaps}. Also, the band gaps of B3LYP, a non-screened hybrid functional, and EXX-OEP, which essentially is a localized version of the pure Hartree-Fock exchange potential, are considerably larger and well within the range of experimental values. A more suitable range separation scheme of the Hartree-Fock exchange might lead to better band gaps for MnO.

However, it is not clear to what extent the transition from the weakly dispersive band at the valence band top to the parabolic band contributes to the low energy optical absorption. In a free ion, a transition from $d$ to $s$ orbitals is symmetry-forbidden, as both intial and end states are even under inversion. This constraint is lifted in the monoxides, where the parabolic $4s$ band has a considerable contribution from Mn $4p$ states away from the $\Gamma$ point. These $p$ states are odd under inversion and thus allow for transitions to the conduction band minimum. At the same time, the valence band top has contributions from O $2p$, which allows for charge transfer from oxygen states to Ni $4s$ states. However, it is possible that these transitions are rather weak due to the mixed nature of the valence band top and the $4s$ band. As for NiO, the predicted energies of the unoccupied $d$ states in sX-LDA are higher than those from HSE03 and similar to those from G$_0$W$_0$@HSE03.

Our cross-weighted DOS nicely reproduces the three dominating peaks found in the experimental XPS spectrum\cite{zimmermann-PESBIS}. The peak at the valence band top arises from $d$ states of $e_g$ character, which are hybridized with O $2p$ states and split from the second, t$_{2g}$ dominated, peak by an energy of 1.8\,eV. This splitting is considerably higher than the splitting in LDA (1.0\,eV) and in excellent agreement with the reported value from scGW of 1.7\,eV and with G$_0$W$_0$@HSE03. Van Elp \textit{et al.}\cite{vanelp-mno} reported a peak splitting of 1.9\,eV in their photoemission experiments. The splitting in EXX-OEP\cite{engel-2009} is approximately 1.4\,eV. The main peak in the conduction band arises from a convolution of weakly dispersive states of mainly Mn $3d$ character at VBM+7\,eV (t$_{2g}$) and VBM+8\,eV (e$_g$) and is predicted to lie slightly higher in energy than the BIS peak by about 0.5\,eV.

%
%
\begin{table}[tb]
\begin{ruledtabular}
\caption{\label{tab:VBW} Valence band widths (in eV) from different theoretical methods compared to experiment.}
\begin{tabular*}{\columnwidth}{@{\extracolsep{\fill}} c  c c c c}
&MnO&FeO&CoO&NiO\\
\hline
sX-LDA&8.25&8.7&8.5&7.5\\
sX-LDA (exp. lat. const.)&7.5&8.3&8.6&8\\
HSE03\cite{roedl-2009}&7.1&8.3&8&7.4\\
EXX-OEP\cite{engel-2009}&$\sim$6&6-8.5&$\sim$8.5&$\sim$8\\
B3LYP&~6.8\cite{feng-2004}&~7.5\cite{alfredsson-2004}&8.2\cite{feng-2004}&~7.5\cite{feng-2004-nio}\\
scGW\cite{faleev-2004}&7.25&&&$\sim$7.8\\
GW$_0$@LDA+U\cite{jiang-tmo}&7.5&9.5&8.5&8\\
G$_0$W$_0$@HSE03\cite{roedl-2009}&7&8.8&8.1&7.4\\
Exp (XES)\cite{kurmaev-2008}&7.5-8&&8-9&8-8.5\\
Exp (UPS)\cite{zimmermann-PESBIS}&7.5-8.5&8-9&8.5-9&8.5-9.5\\
\end{tabular*}
\end{ruledtabular}
\end{table}
\subsubsection{Comparison with XES-XAS spectra}
Lastly, we compare the partial density of states of the oxygen $2p$ electrons with recently reported oxygen x-ray emission spectroscopy and x-ray absorption spectroscopy (XES-XAS) measurements\cite{kurmaev-2008} on MnO, CoO and NiO, in Fig.~\ref{fig:TMO-Ostates}. The shape of the experimental spectra is quite similar for all three materials. The main feature is a sharp peak at 4.0-4.5\,eV below the valence band maximum (VBM), which corresponds to a flat band in the valence band from a mixture of oxygen $2p$ and a metal $3d$ state of $e_g$ symmetry. Three smaller features exist at -1\,eV, -2.3\,eV and -6.5\,eV. Our calculations exhibit all four peaks, albeit at different energies. While the hump from the flat hybrid $2p$/$3d$ band at the valence band maximum is well reproduced by a peak in all our theoretical spectra, the other three features are predicted at lower energies compared to experiments. For MnO, the agreement is quite good, with the dominant peak being shifted to about -5.5\,eV. For CoO and NiO, the peak is being shifted even stronger, to -6.5\,eV and -6\,eV, respectively. The energies of the oxygen $2p$ states are strongly influenced by their hybridization with metal $d$ electrons. The behaviour of these occupied states mirrors our observations for the unfilled states with strong $3d$ contribution. The non-local exchange causes a noticeable down-shift of the occupied $d$ states compared to GGA.\\
\indent Table~\ref{tab:VBW} compares the valence band widths obtained from different theoretical methods with the XES results and ultraviolet photoemission spectroscopy (UPS). Our results from sX-LDA are generally at the upper end of the theoretical values and fit very well to the experimental spectra. The reported HSE03 valence band widths for MnO and NiO agree with our own HSE03 calculations. G$_0$W$_0$@HSE03 inherits the compressed valence bands from HSE03 and introduces only a tiny renormalization. 

\section{Conclusion}
Hybrid functional density functional theory was used to calculate the electronic band structures and density of states of the four transition metal oxides MnO, FeO, CoO and NiO. We conclude that the screened hybrid functional sX-LDA can successfully predict the electronic properties of all four materials with overall similar or greater accuracy than the established hybrid functional HSE03. This confirms that a correct description of the exchange interaction is of foremost priority for these materials. On the other hand, the effect of many-body effects can be reasonably approximated by a suitable screening of the electron exchange. A more sophisticated treatment of correlation effects might lead to even better results. Combined with the capability of self-consistent total energy calculations and relatively low computational cost, this makes screened hybrid functionals interesting alternatives to quasiparticle methods for the simulation of defect properties.



%

\end{document}